\documentclass[a4paper,11pt]{JHEP}
\usepackage[latin1]{inputenc}
\usepackage[T1]{fontenc}
\usepackage{cite}
\usepackage{epsfig}
\usepackage{slashed}
\renewcommand{\d}{\mathrm{d}}
\newcommand{\e}{\mathrm{e}}
\newcommand{\be}{\begin{equation}}
\newcommand{\ee}{\end{equation}}
\newcommand{\co}{\cosh{\mu y}}
\newcommand{\si}{\sinh{\mu y}}

\newcommand{\R}{\mathcal{R}}
\newcommand{\HU}{\mathcal{H}}
\newcommand{\cor}{\cosh{\mu \mathcal{R}}}
\newcommand{\sir}{\sinh{\mu \mathcal{R}}}
\newcommand{\tar}{\tanh{\mu \mathcal{R}}}

\title{Branes on the Horizon}

\author{Anne-Christine Davis\footnote{\texttt{a.c.davis@damtp.cam.ac.uk}}, Christophe Rhodes\footnote{\texttt{c.s.rhodes@damtp.cam.ac.uk}}, Ian Vernon\footnote{\texttt{i.r.vernon@damtp.cam.ac.uk}}\\Department of
  Applied Mathematics and Theoretical Physics, \\Centre for
  Mathematical Sciences, \\Wilberforce Road, University of Cambridge, 
 Cambridge, CB3 0WA, UK}

\abstract{Models with extra dimensions are often invoked to resolve
  cosmological problems. We investigate the possibility of apparent
  acausality as seen by a brane-based observer resulting from signal
  propagation through the extra dimensions. Null geodesics are first
  computed in static and cosmological single-brane models, following
  which we derive the equations of motion for the inter-brane distance
  in a two-brane scenario, which we use to examine possible acausality
  in this more complex setup. Despite observing significant effective
  acausality in some situations there is no \textit{a priori} solution
  to the horizon problem using this mechanism. In the two-brane
  scenario there can be significant late time violation of
  gravitational Lorentz invariance, resulting in the gravitational
  horizon being larger than the particle horizon, leading to potential
  signals in gravitational wave detectors.}

\preprint{DAMTP-2001-44}


\begin{document}

\section{Introduction}

Recently there has been considerable interest in the novel suggestion
that we live in a Universe that possesses more than four
dimensions. The standard model fields  
are assumed to be confined to a hypersurface (or 3-brane)
embedded in this higher dimensional space, in contrast the gravitational
fields propagate through the whole of
spacetime~\cite{ruba,akam1,anton,visser,ahdd,ahddII,aahdd,sund1,rscompact99,rshierarchy99}.
In order for this to be a
phenomenologically relevant model of our universe, 
standard four-dimensional gravity must be recovered on our
brane. There are various ways to do this, the most obvious being to
assume that the extra dimensions transverse to our brane are compact.
In this case gravity can be recovered on scales larger than the size
of the extra dimensions~\cite{ahdd,ahddII,aahdd}. This is different
from earlier proposals since the restrictions on the size of the 
extra dimensions from particle
physics experiments no longer apply, as the standard model fields 
are confined to the brane. The extra dimensions only have to
be smaller than the scale on which gravity experiments have probed,
of order 1mm at the time of writing.
Another way to recover four-dimensional gravity at large distances is 
to embed a positive tension 3-brane into an
AdS$_5$ bulk~\cite{rscompact99,rshierarchy99}. In this scenario four-dimensional
gravity is obtained at scales larger than the AdS radius. Randall
and Sundrum showed that this could produce sensible gravity even if
the extra dimension was not compact. 

The cosmology of these extra dimension scenarios has been investigated
and the Friedman equation derived and shown to contain 
important deviations from the usual 4-dimen\-sional
case~\cite{noncosmo,cosmo1extrad,expans1extrad}. Some inflationary
models have been investigated~\cite{inflation}, as have brane world
phase transitions, topological defects and baryogenesis~\cite{cosmophase}.

The possibility that various cosmological problems could be solved in
an extra-dimen\-sional scenario has been examined by several authors;
however a novel suggestion that could possibly resolve the horizon
problem was made in~\cite{Freese}, where Chung and Freese used a
variety of  simple metrics to  demonstrate that gravity signals,
propagating either purely in the extra dimension or via a second
hidden brane, could connect regions of our four-dimensional universe
naïvely thought to be causally unconnected. Such a mechanism would
have important cosmological consequences as it could greatly alter the
size of the particle horizon as well as possibly providing a method of
experimentally confirming the existence of extra dimensions.

The aim of this present work is to fully investigate both the one and two
brane mechanisms proposed by Chung and Freese, in a much more general,
physically acceptable brane world scenario: that of a non-$Z_2$ symmetric
cosmological model with a non-zero Weyl tensor component
~\cite{tyerscosmo,branevsshell,fpaper,bran1,bran2,bran3}.

In the single brane case we examine numerically the nature of null
geodesics leaving our 3-brane at various times in the universe:
whether they return to our 3-brane or freeze out at the
horizon; their apparent speed relative to that of
light as viewed by a four-dimensional observer; and how these
attributes depend on the initial velocity of the geodesics. The effect
on the geodesics of breaking the $Z_2$ symmetry and of varying the
Weyl tensor component is also investigated fully. We have therefore
considerably generalised the previous work done on the
subject~\cite{geod,shortcuts,csaki}, showing that there are only
significant deviations from Lorentz invariance at early times.

The second case, where signals travel along a hidden brane, has not
previously been investigated in any realistic scenario: the metrics used
in~\cite{Freese} were not solutions to Einstein's equations. In order to
examine the two brane case fully, the non-linear equation of motion of the 
inter-brane distance (otherwise known as the cosmological radion
$\R$~\cite{bin3}) is concisely derived and is used to evaluate the relative
speeds of signals on the second brane compared to the first, quantitatively
for the stationary case and qualitatively for the moving brane case. It is
found that although signals can in some situations propagate along the
second brane significantly faster than along ours, it is not a large
enough effect to solve the horizon problem; on the other hand we do
get significant violation of Lorentz invariance at late times such
that the gravitational horizon is larger than the particle
horizon. This could lead to signals in future gravitational wave
detector experiments.

The paper is organised as follows: the general setup and geodesic
equations are derived in section \ref{sec:setup}); section
\ref{sec:RS} contains a brief discussion of geodesics in the normal
Randall--Sundrum model \cite{geod}. We start, in section \ref{sec:static}, by investigating non-cosmological (i.e.\ static)
solutions of Einstein's equations, with the Randall--Sundrum model
extended by admitting \cite{fpaper} the possibility of $Z_2$-symmetry
breaking across the brane, as well as perturbations from perfect
tuning; the behaviour of null geodesics in the five-dimensional theory
is examined both analytically and numerically. We then extend the
investigation in section \ref{sec:cosmo} to fully cosmological
solutions of Einstein's equations without mirror symmetry, numerically
investigating the possibilities for solving the horizon problem in a
similar fashion to \cite{csaki}. In section \ref{sec:twobrane} we
consider the possible acausality in a two brane scenario where in
general the second brane is moving. Our conclusions are discussed in
section \ref{sec:conclusions}.

\section{Geodesics in 5-dimensional Einstein Gravity}
\label{sec:setup}

Throughout this paper we investigate the behaviour of geodesics in a
variety of different brane world scenarios. In this section we
describe both the general setup and the assumptions that have been
made, as well as giving the equations from which all the results in
this paper are derived.

In each scenario that we investigate, it is assumed that the
five-dimensional spacetime satisfies three-dimensional homogeneity and
isotropy as we are mainly interested in realistic cosmological
models. We choose to work in a `brane-based' coördinate system as it
will facilitate physical interpretation of the results. We therefore assume 
that the four-dimensional universe in which we
reside is situated on a 3-brane which is chosen to be at the origin of
the extra dimension ($y=0$). This implies that the most general metric
must have the form
\begin{equation}\label{metAnz}
  \d s^2 = - n^2(t,y)\d t^2 + a^2(t,y)\gamma_{ij}\d x^i \d x^j + \d y^2,
\end{equation}
where $t$ is the cosmic time on our brane, $x^i$ represents the three
spatial dimensions of our brane, and $y$ is the coördinate of the
extra dimension. $\gamma_{ij}$ is the maximally symmetric
three-dimensional metric with $k = -1,0,1$ parameterizing the spatial
curvature, although throughout most of this paper $k$ will be set to
zero. The bulk is assumed to be empty except for a
cosmological constant $\Lambda$, and therefore the metric is obtained
by solving the five-dimensional Einstein's equations\footnote{We use
the standard brane world convention in that
lower-case Roman indices (such as $i$, $j$) run across the normal 
space dimensions (1 to 3); Greek indices ($\mu$, $\nu$) run across 
time and normal space (0 to 3), while capital Roman indices 
($A$, $B$) cover all space and time dimensions (0 to 3 and 5).},
\begin{equation}
G_{AB} \; - \; \Lambda g_{AB} \;\; = \;\; \kappa^2 T_{AB},
\end{equation}
where we define $\kappa^2 \equiv 1/\widetilde{M}_5^3$,
$\widetilde{M}_5$ being the fundamental (reduced) five-dimensional Planck Mass. In
the single brane scenario, the energy-momentum tensor $T_{AB}$ takes
the form,
\begin{equation}
T_{AB} \;\; = \;\; \delta(y) \mathrm{diag}(\rho_0,P_0,P_0,P_0,0),
\end{equation}
and for the two brane scenario,
\begin{equation}
T_{AB} \;\; = \;\; \delta(y) \mathrm{diag}(-\rho_0,P_0,P_0,P_0,0) \;+\;
               \delta(y-\R(t)) \mathrm{diag}(-\rho_2,P_2,P_2,P_2,0).
\end{equation}
where we follow the notation used in~\cite{bin3} by defining $\rho_0$
and $\rho_2$ to be the energy density of our brane and of the 
second brane respectively, and by defining the pressures $P_0$ and
$P_2$ similarly. The position of the second brane given by $y=\R(t)$, is in
general time dependent. 

In Sections (\ref{sec:static}) and (\ref{sec:cosmo}) we use solutions
of Einstein's equations that do not possess a $Z_2$ symmetry (or
mirror symmetry) across the brane. For a full discussion of this topic
see~\cite{tyerscosmo,branevsshell,fpaper,bran1,bran2,bran3}.

\subsection{Geodesic Equations}

Here we derive the necessary geodesic equations corresponding to the 
metric given by equation (\ref{metAnz}). In what follows the notation
is such that dots indicate differentiation with
respect to the affine parameter and dashes with respect to the fifth
dimension, $y$, leaving $\frac{\partial}{\partial t}$ as the differentiation
operator with respect to coördinate time. Starting from the variational
principle
\begin{equation}
  \label{Ldens}
  \delta S = \delta\! \int\!\! \mathcal{L} \d s = 0,
\end{equation}
with
\begin{equation}
  \mathcal{L} = \sqrt{g_{AB}       \dot{x}^A\dot{x}^B},
\end{equation}
we can derive the equations of motion for test particles. We shall
principally be concerned here with lightlike geodesics, which means
that we can consider the variation of $\int \! \mathcal{L}^2 \d s$,
so that the effective Lagrangian density is
$-n^2(t,y)\dot{t}^2+a^2(t,y)\dot{x}_i\dot{x}_i + \dot{y}^2$, giving
Euler-Lagrange equations:
\begin{eqnarray}
  \nonumber \ddot{t}&=&\frac{1}{n}\left(\frac{\partial n}{\partial t}\dot{t}^2 -
    2\dot{n}\dot{t} - \frac{\partial a}{\partial t} \frac{\theta_i
      \theta_i}{a^3n}\right)\\
  \dot{x}_i&=&\frac{\theta_i}{a^2}\\
  \nonumber \ddot{y} &=& -\frac{\theta_i \theta_i}{a^2}\left(\frac{n'}{n}-\frac{a'}{a}\right)-\frac{n'}{n}\dot{y}^2,
\end{eqnarray}
where $\theta_i$ are integration constants. For null geodesics,
with which we shall be exclusively concerned
in this study, we have the additional constraint that 
the test particle must move at the speed of light; this tells us that
the first integral of the $t$ equation above must be of the form
\begin{equation}
n^2\dot{t}^2 = \dot{y}^2 + \frac{\theta_i \theta_i}{a^2}.
\end{equation}
These equations will be used  extensively throughout
sections \ref{sec:RS}, \ref{sec:static} and \ref{sec:cosmo} to 
evaluate possible shortcuts through the bulk.

\section{Geodesics in the Randall-Sundrum Model}
\label{sec:RS}

\subsection{The Randall-Sundrum Metric}

The Randall-Sundrum model, as initially presented
\cite{rscompact99,rshierarchy99} is not a cosmological one; rather, it
is included here as a simple model to develop the reasoning employed
in rather more complicated cases.  It may be derived from the above,
general, metric Ansatz given in equation (\ref{metAnz}) by  taking
$n(\slashed{t},y) = a(\slashed{t},y) = a(y)$, and then solving
Einstein's equations while assuming a $Z_2$ symmetry across each
brane. Provided that the bulk cosmological constant $\Lambda$ is less
than zero, and that energy densities of each of the branes are tuned
such that
\begin{equation}
\kappa^2 \rho_0 \; = \; -\kappa^2 \rho_2 \; = \; 
                        \sqrt{-6\Lambda},
\end{equation}
the familiar Randall-Sundrum metric is recovered:
\begin{equation}
\d s^2 = \e^{-2\mu|y|}\eta_{\mu\nu}\d x^\mu \d x^\nu + \d y^2,
\end{equation}
which implies that
\begin{equation}
n(y) = a(y) = \e^{-\mu|y|}.
\end{equation}
$\mu$ is the inverse of the AdS$_5$ curvature radius and is given by
$\mu = \sqrt{-\Lambda/6}$. 

\subsection{RS Geodesics}

We can immediately see that the Randall-Sundrum model is not going to
help us solve the horizon problem. The local speed of light is
everywhere the same, because $n(y) = a(y)$, and so even if off-brane
null geodesics can return to the brane, they will return with
an effective sub-luminal velocity.

However, the Randall-Sundrum model does not exhibit even this kind of
behaviour; as shown in \cite{geod}, the geodesic follows the path
given by
\begin{equation}
  \e^{2\mu y} = \e^{2\mu y_0} + 2\mu\dot{y}_0\e^{2\mu y_0}t - v^2\mu^2t^2,
\end{equation}
with $v^2 = \eta_{\mu\nu}\dot{x}^\mu\dot{x}^\nu$. 

As a side note, $v^2 = 0$ does not necessarily correspond to a
lightlike geodesic, as claimed in section 2 of \cite{geod}, as the
condition for lightlike behaviour is $g_{AB}\dot{x}^A\dot{x}^B = 0$,
which includes an extra $\dot{y}^2$.

\section{Geodesics in Other Static Braneworld Models}
\label{sec:static}

\subsection{Obtaining the Metric}

We relax some of the assumptions made in the Randall-Sundrum model;
specifically, we no longer impose strict $Z_2$ symmetry, nor do we
fine-tune the bulk cosmological constant. Further to the relaxation of
fine-tuning and $Z_2$ symmetry, we must also adapt the Ansatz so that
$n(y)$ is no longer equal to $a(y)$ to obtain a self-consistent
solution.

The $G_{00}$ equation from \cite{noncosmo} yields
\begin{equation}
a^2(y) = \cosh(2\mu y) + A\sinh(2\mu y),
\end{equation}
where we have implicitly scaled so that $a_0 = a|_{y=0} = 1$. To
consider the loss of $Z_2$ or mirror symmetry, when we apply the
Israel junction condition
\begin{equation}
[a'(y)] = -\frac{\kappa^2}{3}\rho,
\end{equation}
where $[f] = f(0^{+})-f(0^{-})$, we make the assumption that
$a'(0^{+}) = -a'(0^{-}) + d_a$, giving
\begin{equation}
\mu A = -\frac{\kappa^2}{6} \rho + \frac{d_a}{2};
\end{equation}
$d_a = 0$ for $Z_2$-symmetric braneworlds. Then, parameterizing the
lack of fine-tuning by
\begin{equation}
r = \frac{\kappa^2\rho}{6\mu}
\end{equation}
and the $Z_2$ symmetry breaking by
\begin{equation}
d = \frac{d_a}{2\mu},
\end{equation}
we obtain
\begin{equation}
  \label{eq:statica}
  a(y) = \sqrt{\cosh(2\mu|y|) + (d - r) \sinh(2\mu|y|)}.
\end{equation}
The $G_{55}$ equation then yields
\begin{equation}
  \label{eq:staticn}
  n(y) = \frac{\sinh(2\mu|y|) + (d - r) \cosh(2\mu|y|)}
{(d-r)\sqrt{\cosh(2\mu|y|) + (d - r) \sinh(2\mu|y|)}},
\end{equation}
which has again been scaled so that $n(y=0)=1$; this solution may then
be substituted into the other Einstein equations and junction
conditions to check that it is self-consistent.

\subsection{Geodesics in Static Non-$Z_2$ Brane Worlds}

Since the Lagrangian density
$\mathcal{L}$ given by equation (\ref{Ldens}) does not depend explicitly on
coördinate time $t$, we may obtain a further first integral of the
form
\begin{equation}
  \dot{t} = \frac{\theta_0}{n^2(y)},
\end{equation}
and so straight from the Lagrangian constraint for lightlike geodesics
we have
\begin{equation}
  \dot{y}^2 = \frac{\theta_0^2}{n^2(y)} - \frac{\theta_i\theta_i}{a(y)^2}.
\end{equation}
This is effectively a conservation of energy equation, with a kinetic
term proportional to $\dot{y}^2$ and a potential of the form
\begin{equation}
  V(y) = -\frac{\theta_0^2}{n^2(y)} + \frac{\theta_i\theta_i}{a(y)^2}.
\end{equation}
We may therefore reason that firstly if a geodesic starts its life on our
brane (at $y=0$) then we have, since $\dot{y}^2 \geq 0$,
\begin{equation}
  \theta_0^2 \geq \theta_i\theta_i.
\end{equation}
Further, if the geodesic is ever to return, then $\dot{y}$ must at
some point be zero, so
\begin{equation}
  \theta_0^2\left(\frac{1}{n^2(y)} - \frac{\gamma^2}{a^2(y)}\right),
\end{equation}
where $0 < \gamma^2 = \frac{\theta_i\theta_i}{\theta_0^2} < 1$, must be zero
at some positive value of $y$. Rearranging and substituting the above
solution for the metric, we find that the positions at which
$\dot{y}=0$, denoted by $y^{\pm}_t$, satisfies
\begin{equation}
  \cosh(2\mu y^{\pm}_t) + (d-r)\sinh(2\mu y^{\pm}_t) = 
\pm \gamma \frac{\sinh(2\mu y^{\pm}_t) +
    (d-r)\cosh(2\mu y^{\pm}_t)}{(d-r)},
\end{equation}
or
\begin{equation}
  \label{eq:staticturn}
  y^{\pm}_t = \frac{1}{2\mu}\tanh^{-1}\left[\frac{(d-r)(\pm \gamma-1)}{(d-r)^2
  \mp \gamma}\right].
\end{equation}
However, there could be a horizon intervening between $y=0$ and this
point; a horizon will occur if either of $a(y)$ or $n(y)$ is equal to
zero. We split the analysis into two parts, one for $(d-r)> -1$ and
one for $(d-r)<-1$, reminding the reader that $(d-r)=-1$ corresponds to
the Randall-Sundrum braneworld. Below, subscript $t$ refers to the
geodesic turning point, while subscript $h$ to the position of the
horizon.

\subsubsection{The $(d-r)>-1$ Case}

\begin{figure}
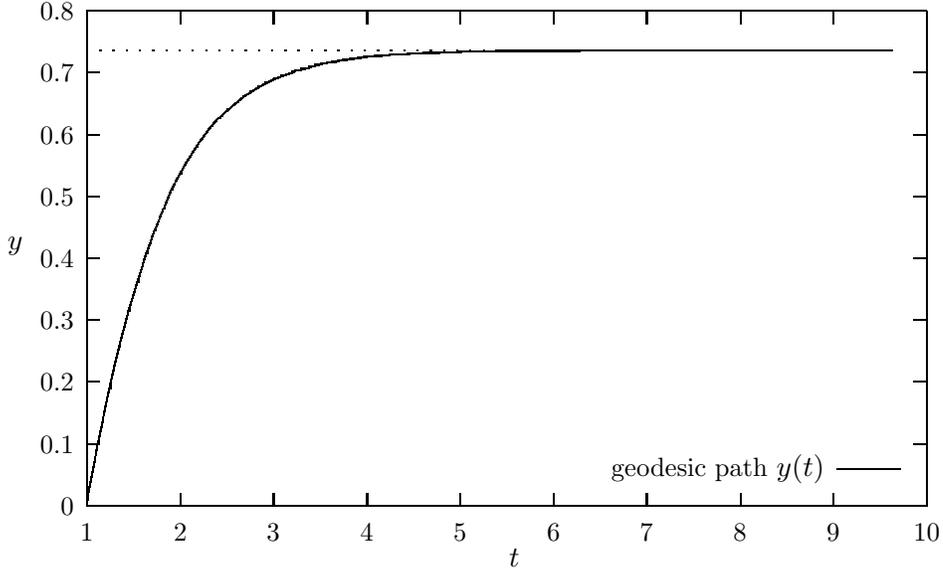

  \begin{center}
    \include{baz}
    \caption{Graph of $y$ against coördinate time for a geodesic in a
      static untuned ($d-r=-0.9$) braneworld with $\dot{x}_0 = -0.5$. The
      dotted line represents the calculated $y^>_h$ horizon position.}
    \label{baz}
  \end{center}
\end{figure}

\begin{figure}
  \begin{center}
\setlength{\unitlength}{0.240900pt}
\ifx\plotpoint\undefined\newsavebox{\plotpoint}\fi
\begin{picture}(1500,900)(0,0)
\font\gnuplot=cmr10 at 10pt
\gnuplot
\sbox{\plotpoint}{\rule[-0.200pt]{0.400pt}{0.400pt}}%
\put(160.0,82.0){\rule[-0.200pt]{4.818pt}{0.400pt}}
\put(140,82){\makebox(0,0)[r]{-0.35}}
\put(1419.0,82.0){\rule[-0.200pt]{4.818pt}{0.400pt}}
\put(160.0,193.0){\rule[-0.200pt]{4.818pt}{0.400pt}}
\put(140,193){\makebox(0,0)[r]{-0.3}}
\put(1419.0,193.0){\rule[-0.200pt]{4.818pt}{0.400pt}}
\put(160.0,304.0){\rule[-0.200pt]{4.818pt}{0.400pt}}
\put(140,304){\makebox(0,0)[r]{-0.25}}
\put(1419.0,304.0){\rule[-0.200pt]{4.818pt}{0.400pt}}
\put(160.0,415.0){\rule[-0.200pt]{4.818pt}{0.400pt}}
\put(140,415){\makebox(0,0)[r]{-0.2}}
\put(1419.0,415.0){\rule[-0.200pt]{4.818pt}{0.400pt}}
\put(160.0,527.0){\rule[-0.200pt]{4.818pt}{0.400pt}}
\put(140,527){\makebox(0,0)[r]{-0.15}}
\put(1419.0,527.0){\rule[-0.200pt]{4.818pt}{0.400pt}}
\put(160.0,638.0){\rule[-0.200pt]{4.818pt}{0.400pt}}
\put(140,638){\makebox(0,0)[r]{-0.1}}
\put(1419.0,638.0){\rule[-0.200pt]{4.818pt}{0.400pt}}
\put(160.0,749.0){\rule[-0.200pt]{4.818pt}{0.400pt}}
\put(140,749){\makebox(0,0)[r]{-0.05}}
\put(1419.0,749.0){\rule[-0.200pt]{4.818pt}{0.400pt}}
\put(160.0,860.0){\rule[-0.200pt]{4.818pt}{0.400pt}}
\put(140,860){\makebox(0,0)[r]{0}}
\put(1419.0,860.0){\rule[-0.200pt]{4.818pt}{0.400pt}}
\put(160.0,82.0){\rule[-0.200pt]{0.400pt}{4.818pt}}
\put(160,41){\makebox(0,0){1}}
\put(160.0,840.0){\rule[-0.200pt]{0.400pt}{4.818pt}}
\put(302.0,82.0){\rule[-0.200pt]{0.400pt}{4.818pt}}
\put(302,41){\makebox(0,0){2}}
\put(302.0,840.0){\rule[-0.200pt]{0.400pt}{4.818pt}}
\put(444.0,82.0){\rule[-0.200pt]{0.400pt}{4.818pt}}
\put(444,41){\makebox(0,0){3}}
\put(444.0,840.0){\rule[-0.200pt]{0.400pt}{4.818pt}}
\put(586.0,82.0){\rule[-0.200pt]{0.400pt}{4.818pt}}
\put(586,41){\makebox(0,0){4}}
\put(586.0,840.0){\rule[-0.200pt]{0.400pt}{4.818pt}}
\put(728.0,82.0){\rule[-0.200pt]{0.400pt}{4.818pt}}
\put(728,41){\makebox(0,0){5}}
\put(728.0,840.0){\rule[-0.200pt]{0.400pt}{4.818pt}}
\put(871.0,82.0){\rule[-0.200pt]{0.400pt}{4.818pt}}
\put(871,41){\makebox(0,0){6}}
\put(871.0,840.0){\rule[-0.200pt]{0.400pt}{4.818pt}}
\put(1013.0,82.0){\rule[-0.200pt]{0.400pt}{4.818pt}}
\put(1013,41){\makebox(0,0){7}}
\put(1013.0,840.0){\rule[-0.200pt]{0.400pt}{4.818pt}}
\put(1155.0,82.0){\rule[-0.200pt]{0.400pt}{4.818pt}}
\put(1155,41){\makebox(0,0){8}}
\put(1155.0,840.0){\rule[-0.200pt]{0.400pt}{4.818pt}}
\put(1297.0,82.0){\rule[-0.200pt]{0.400pt}{4.818pt}}
\put(1297,41){\makebox(0,0){9}}
\put(1297.0,840.0){\rule[-0.200pt]{0.400pt}{4.818pt}}
\put(1439.0,82.0){\rule[-0.200pt]{0.400pt}{4.818pt}}
\put(1439,41){\makebox(0,0){10}}
\put(1439.0,840.0){\rule[-0.200pt]{0.400pt}{4.818pt}}
\put(160.0,82.0){\rule[-0.200pt]{308.111pt}{0.400pt}}
\put(1439.0,82.0){\rule[-0.200pt]{0.400pt}{187.420pt}}
\put(160.0,860.0){\rule[-0.200pt]{308.111pt}{0.400pt}}
\put(160.0,82.0){\rule[-0.200pt]{0.400pt}{187.420pt}}
\put(1279,820){\makebox(0,0)[r]{geodesic path $x(t)$}}
\put(20,491){\makebox(0,0){$x$}}
\put(800,1){\makebox(0,0){$t$}}
\put(1299.0,820.0){\rule[-0.200pt]{24.090pt}{0.400pt}}
\put(160,860){\usebox{\plotpoint}}
\put(160.17,842){\rule{0.400pt}{3.300pt}}
\multiput(159.17,851.15)(2.000,-9.151){2}{\rule{0.400pt}{1.650pt}}
\multiput(162.58,829.34)(0.494,-3.762){29}{\rule{0.119pt}{3.050pt}}
\multiput(161.17,835.67)(16.000,-111.670){2}{\rule{0.400pt}{1.525pt}}
\multiput(178.58,712.79)(0.495,-3.306){33}{\rule{0.119pt}{2.700pt}}
\multiput(177.17,718.40)(18.000,-111.396){2}{\rule{0.400pt}{1.350pt}}
\multiput(196.58,597.40)(0.494,-2.823){27}{\rule{0.119pt}{2.313pt}}
\multiput(195.17,602.20)(15.000,-78.199){2}{\rule{0.400pt}{1.157pt}}
\multiput(211.58,515.16)(0.494,-2.590){25}{\rule{0.119pt}{2.129pt}}
\multiput(210.17,519.58)(14.000,-66.582){2}{\rule{0.400pt}{1.064pt}}
\multiput(225.58,445.69)(0.493,-2.122){23}{\rule{0.119pt}{1.762pt}}
\multiput(224.17,449.34)(13.000,-50.344){2}{\rule{0.400pt}{0.881pt}}
\multiput(238.58,392.89)(0.494,-1.745){25}{\rule{0.119pt}{1.471pt}}
\multiput(237.17,395.95)(14.000,-44.946){2}{\rule{0.400pt}{0.736pt}}
\multiput(252.58,345.86)(0.493,-1.448){23}{\rule{0.119pt}{1.238pt}}
\multiput(251.17,348.43)(13.000,-34.430){2}{\rule{0.400pt}{0.619pt}}
\multiput(265.58,309.63)(0.493,-1.210){23}{\rule{0.119pt}{1.054pt}}
\multiput(264.17,311.81)(13.000,-28.813){2}{\rule{0.400pt}{0.527pt}}
\multiput(278.58,279.26)(0.492,-1.013){21}{\rule{0.119pt}{0.900pt}}
\multiput(277.17,281.13)(12.000,-22.132){2}{\rule{0.400pt}{0.450pt}}
\multiput(290.58,256.16)(0.493,-0.734){23}{\rule{0.119pt}{0.685pt}}
\multiput(289.17,257.58)(13.000,-17.579){2}{\rule{0.400pt}{0.342pt}}
\multiput(303.58,237.51)(0.492,-0.625){21}{\rule{0.119pt}{0.600pt}}
\multiput(302.17,238.75)(12.000,-13.755){2}{\rule{0.400pt}{0.300pt}}
\multiput(315.00,223.92)(0.496,-0.492){21}{\rule{0.500pt}{0.119pt}}
\multiput(315.00,224.17)(10.962,-12.000){2}{\rule{0.250pt}{0.400pt}}
\multiput(327.00,211.93)(0.669,-0.489){15}{\rule{0.633pt}{0.118pt}}
\multiput(327.00,212.17)(10.685,-9.000){2}{\rule{0.317pt}{0.400pt}}
\multiput(339.00,202.93)(0.758,-0.488){13}{\rule{0.700pt}{0.117pt}}
\multiput(339.00,203.17)(10.547,-8.000){2}{\rule{0.350pt}{0.400pt}}
\multiput(351.00,194.93)(1.033,-0.482){9}{\rule{0.900pt}{0.116pt}}
\multiput(351.00,195.17)(10.132,-6.000){2}{\rule{0.450pt}{0.400pt}}
\multiput(363.00,188.94)(1.651,-0.468){5}{\rule{1.300pt}{0.113pt}}
\multiput(363.00,189.17)(9.302,-4.000){2}{\rule{0.650pt}{0.400pt}}
\multiput(375.00,184.94)(1.651,-0.468){5}{\rule{1.300pt}{0.113pt}}
\multiput(375.00,185.17)(9.302,-4.000){2}{\rule{0.650pt}{0.400pt}}
\multiput(387.00,180.95)(2.472,-0.447){3}{\rule{1.700pt}{0.108pt}}
\multiput(387.00,181.17)(8.472,-3.000){2}{\rule{0.850pt}{0.400pt}}
\put(399,177.17){\rule{2.300pt}{0.400pt}}
\multiput(399.00,178.17)(6.226,-2.000){2}{\rule{1.150pt}{0.400pt}}
\put(410,175.17){\rule{2.500pt}{0.400pt}}
\multiput(410.00,176.17)(6.811,-2.000){2}{\rule{1.250pt}{0.400pt}}
\put(422,173.67){\rule{2.891pt}{0.400pt}}
\multiput(422.00,174.17)(6.000,-1.000){2}{\rule{1.445pt}{0.400pt}}
\put(434,172.67){\rule{3.132pt}{0.400pt}}
\multiput(434.00,173.17)(6.500,-1.000){2}{\rule{1.566pt}{0.400pt}}
\put(447,171.67){\rule{2.650pt}{0.400pt}}
\multiput(447.00,172.17)(5.500,-1.000){2}{\rule{1.325pt}{0.400pt}}
\put(458,170.67){\rule{3.854pt}{0.400pt}}
\multiput(458.00,171.17)(8.000,-1.000){2}{\rule{1.927pt}{0.400pt}}
\put(160.0,858.0){\rule[-0.200pt]{0.400pt}{0.482pt}}
\put(489,169.67){\rule{2.650pt}{0.400pt}}
\multiput(489.00,170.17)(5.500,-1.000){2}{\rule{1.325pt}{0.400pt}}
\put(474.0,171.0){\rule[-0.200pt]{3.613pt}{0.400pt}}
\put(537,168.67){\rule{2.168pt}{0.400pt}}
\multiput(537.00,169.17)(4.500,-1.000){2}{\rule{1.084pt}{0.400pt}}
\put(500.0,170.0){\rule[-0.200pt]{8.913pt}{0.400pt}}
\put(546.0,169.0){\rule[-0.200pt]{202.356pt}{0.400pt}}
\end{picture}
    \caption{Graph of $x$ against coördinate time for a geodesic in a
      static untuned ($d-r=0.9$) braneworld with $\dot{x}_0 =
      -0.5$. Note that as the geodesic goes towards the horizon its
      $x$-velocity decreases.}
    \label{quux}
  \end{center}
\end{figure}
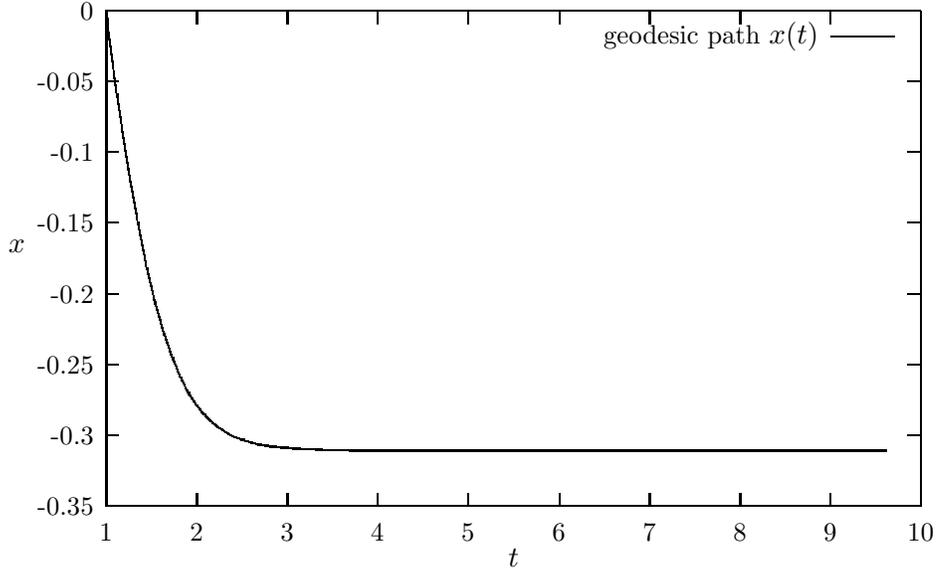

For $(d-r)>-1$, $a(y)$ (equation \ref{eq:statica}) is never zero
because $\cosh(2\mu y)>\sinh(2\mu y)$ for all $y$. $n(y)=0$ gives
(from equation \ref{eq:staticn}) 
\begin{equation}
  y^>_h = \frac{1}{2\mu}\tanh^{-1}(r-d).
\end{equation}
Then $y^-_t$ (equation \ref{eq:staticturn}) is clearly greater than
$y^>_h$ (as $(d-r)^2 < 1$); also $y^+_t$ is either greater than
$y^>_h$ or is negative (the latter when $(d-r)^2<\gamma$).  Thus a
geodesic leaving our brane in the positive $y$ direction for
$(d-r)>-1$ will always reach a horizon at $y = y^>_h$, and will for an
observer on the brane therefore never return. See figures \ref{baz}
and \ref{quux} for illustrations of a sample geodesic in this
régime.

\subsubsection{The $(d-r)<-1$ Case}

\begin{figure}
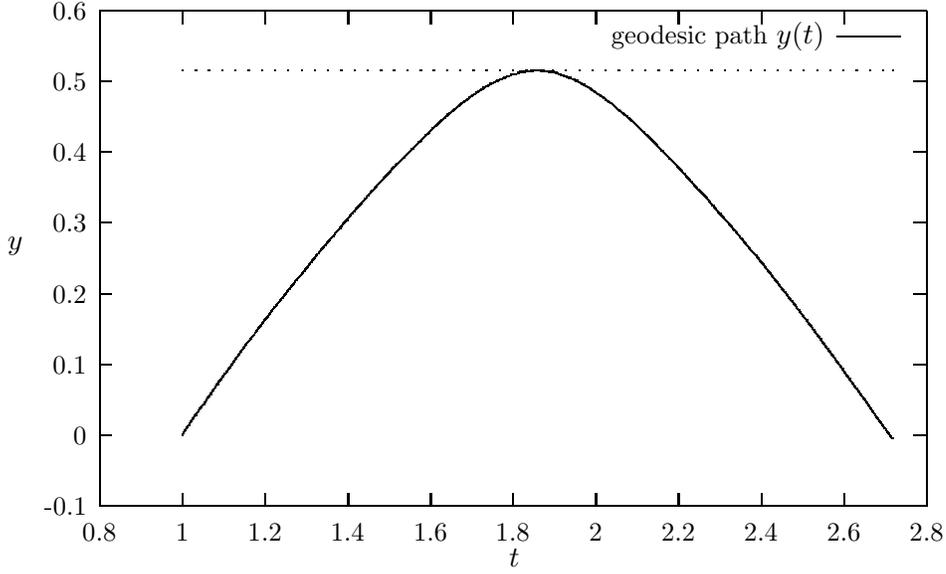

  \begin{center}
    \include{foo}
    \caption{Graph of $y$ against coördinate time for a geodesic in a
      static untuned ($d-r=-1.1$) braneworld with $\dot{x}_0 = -0.5$. The
      dotted line represents the calculated $y$ turnaround position.}
    \label{foo}
  \end{center}
\end{figure}

\begin{figure}
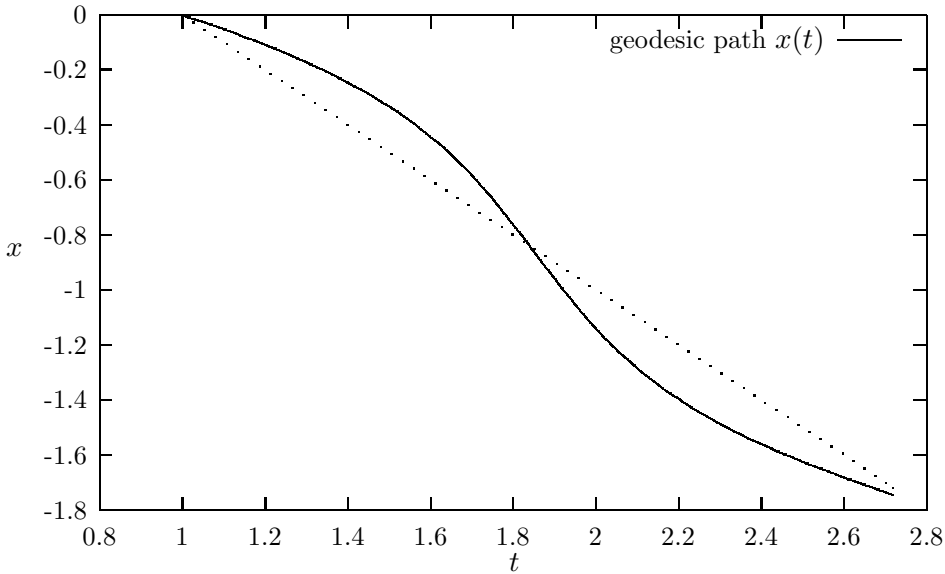

  \begin{center}
    \include{bar}
    \caption{Graph of $x$ against coördinate time for a geodesic in a
      static untuned ($d-r=-1.1$) braneworld with $\dot{x}_0 = -0.5$.
      The dotted line represents a light ray moving on the brane. Note
      that the true null geodesic has covered a greater distance on
      return to the brane than the light ray in the same time $t$;
      lower magnitudes of $\dot{x}_0$ lead to higher
      superluminosities.}
    \label{bar}
  \end{center}
\end{figure}

For $(d-r)<-1$, $n(y)$ is never zero, and $a(y) = 0$ gives 
\begin{equation}
  y^<_h = \frac{1}{2\mu}\tanh^{-1}\frac{1}{(r-d)}. 
\end{equation}
This is always greater than
$y^-_t$ which is the smaller of the two turning points, and so for 
$(d-r) < -1$, the geodesic will turn around at a $y$ value of
\begin{equation}
  y_t = \frac{1}{2\mu}\tanh^{-1}\left[\frac{(r-d)(1-\gamma)}{(d-r)^2 - \gamma}\right].
\end{equation}
Figures \ref{foo} and \ref{bar} similarly show a sample geodesic for
$(d-r)<-1$.

\section{Cosmological Models}
\label{sec:cosmo}

\subsection{The non-$Z_2$ Symmetric Cosmological Metric}
In this Section we investigate the behaviour of geodesics in the full
cosmologically realistic, non-$Z_2$ symmetric brane world scenario
(see~\cite{fpaper} for more details). 
In order to obtain standard cosmology at late times we make the usual
assumption, first noted by~\cite{cosmo1extrad}, that our brane 
possesses an energy density which is the
sum of the brane tension $\sigma$ and a physical energy
density $\rho$, and we assume that the brane tension and bulk
cosmological constant are tuned to ensure that the effective
four-dimensional cosmological constant is zero. The Friedman equation
is then given by~\cite{fpaper},
\begin{equation}\label{Freid1}
H_0^2\;\; = \;\;\left( 
      \frac{\dot{a}_0}{a_0}
\right)^2 \;\; = \;\;
\frac{\kappa^4 \sigma}{18} \rho  \; + \; 
\frac{\kappa^4}{36}\rho^2 \; - \; \frac{k}{a_0^2} \; + \; 
\frac{\mathcal{C}}{a_0^4} \; + \; \frac{F^2}{(\rho + \sigma)^2 a_0^8},
\label{eq:friedman}
\end{equation}
where $a_0(t) = a(t,0)$, $\mathcal{C}$ is the dark radiation term
corresponding to a non-zero Weyl tensor component, and $F$ represents
the extent to which the $Z_2$ symmetry is broken. Motivated by
previous work on geodesics~\cite{shortcuts} which suggests that more
interesting effects are to be seen at earlier times, we assume a
radiation dominated universe and therefore set 
\begin{equation}
  \rho = \frac{\lambda}{a_0(t)^4},
\end{equation}
where $\lambda$ is some constant.
We can now use the bulk solutions to the metric Ansatz (\ref{metAnz}) 
which have been
found for a single brane without
reflection symmetry in an infinite fifth dimension~\cite{fpaper}, 
and are given by 
\begin{equation}\label{met1_form}
      a^2(t,y)  =  a^2_0(t) \left( 
                           A(t)\cosh 2\mu y + B(t)\sinh 2\mu y + C(t)
                           \right), 
  \end{equation}
where $\mu$ as before, is defined in terms of the bulk
cosmological constant $\Lambda$ as $\mu = \sqrt{-\Lambda/6}$. The
purely time dependent constants $A(t)$, $B(t)$, and $C(t)$ are given by
\begin{eqnarray} \label{our_soln}
  A(t) \;\;& = &\;\; 1 \;+\; (1+w) \frac{\rho}{\sigma} \;+\; 
                \frac{1}{2}\frac{\rho^2}{\sigma^2} \;+\;
                \frac{f^2 \rho^2}{2(\sigma + \rho)^2}  \\
  B(t) \;\;& = &\;\; -(1+\frac{\rho}{\sigma}) \;\pm\; 
                \frac{f \rho}{(\sigma + \rho)}   \\
  C(t) \;\;& = &\;\; -(1+w)\frac{\rho}{\sigma} \;-\; 
                \frac{1}{2}\frac{\rho^2}{\sigma} \;-\;
                \frac{f^2 \rho^2}{2(\sigma + \rho)^2}  .  
\end{eqnarray}
Here $w$ and $f$ are dimensionless constants defined by 
$w \equiv 18\mathcal{C}/\kappa\sigma\lambda$ and $f\equiv
6F/\kappa^2\sigma \lambda$, and the $\pm$-signs in the expression for
$B(t)$ give the two different solutions on either side of the brane.
The solution for $n(t,y)$ is given in terms of $a(t,y)$ as:
\begin{equation}\label{def_n}
n(t,y) \;\;\; = \;\;\; \frac{\dot{a}(t,y)}{\dot{a}_0(t)},
\end{equation}
where from now the dots denote differentiation with respect to
coördinate time.

To find $a_0(t)$, we must solve the Friedman equation
(\ref{Freid1}); however, for $f \neq 0$, there is no analytical
solution, and so it must be solved numerically; this presents a
problem, however, due to the initial singularity. Therefore, to solve
this equation we approximate the $f^2$ term for early times, when
$\rho\gg\sigma$, as $f^2\sigma^2$; the solution is then, as in
\cite{fpaper},
\begin{equation}
a_0 = \left\{\gamma f
  \left(\frac{1+w}{f\sigma}\left[\cosh\frac{2\sigma f}{3\tilde{M}_5^3}t-1\right]+
\frac{1}{\sigma}\sinh\frac{2\sigma f}{3\tilde{M}_5^3}t\right)\right\}^\frac{1}{4}.
\end{equation}
This solution will be approximately valid to some time $t_0$; as an
estimate of that time, we find when $\rho = 4\sigma$, giving
\begin{equation}
\frac{1+w}{f}\left[\cosh\frac{2\sigma f}
{3\tilde{M}_5^3}t_0-1\right]+\sinh\frac{2\sigma f}{3\tilde{M}_5^3}t_0=\frac{f}{4},
\end{equation}
or
\begin{equation}
t_0 = \frac{3\tilde{M}_5^3}{2\sigma f}\log\left\{\frac{\frac{1+w}{f}+
\frac{f}{4}+\sqrt{(\frac{1+w}{f}+\frac{f}{4})^2-(\frac{1+w}{f})^2+1}}
{\frac{1+w}{f}+1}\right\}.
\end{equation}

\begin{figure}
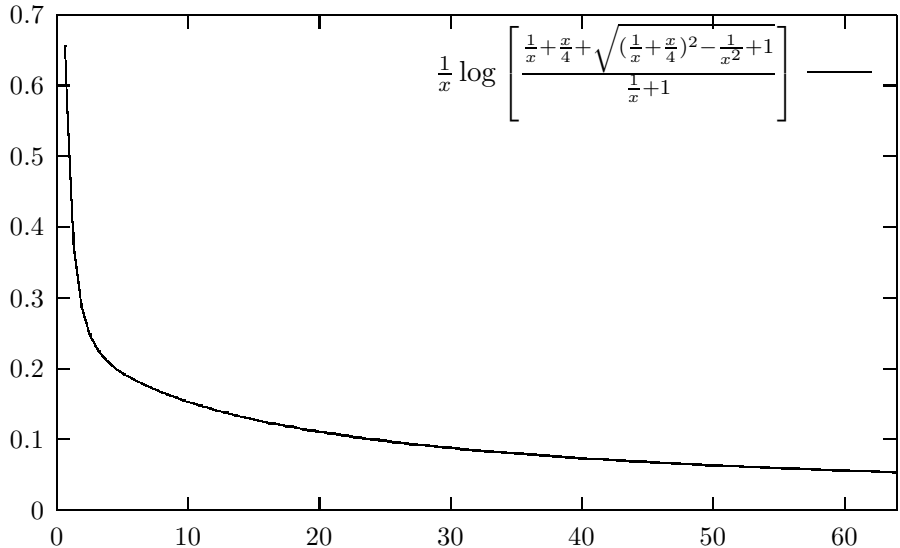

  \begin{center}
    \include{f.t0}
    \caption{A graph of $t_0$ (for $w = 0$) over a range of values of
      $f$; this is used to determine the point at which numerical
      integration of the Friedman equation must begin.}
    \label{f.t0}
  \end{center}
\end{figure}

Over the range of values of interest ($0<w<0.2$, $0<f<60$), $t_0 >
\frac{1}{20}\frac{3\tilde{M}_5^3}{2\sigma f}$, (see figure
\ref{f.t0}) and so we may safely take the initial conditions for
numerical integration of the Friedman equation as our approximate
solution at $t = \frac{1}{40}\frac{3\tilde{M}_5^3}{2\sigma f}$.
This can then be numerically integrated using the full Friedman
equation (\ref{eq:friedman}) to generate the solutions of the Einstein
equations.

\subsection{Geodesics}

We would expect at late times for geodesics to be similar to
the static case, assuming that the dynamical timescale is much smaller
than the cosmological one. This appears broadly to be the case, though
the position or cross-section of the horizon would need to be tuned
for an adequate comparison. The main difference is that for large
dynamical times there will be a certain amount of asymmetry in the
geodesic's path, due to the changing scale factor $a = a(t,y)$.

\begin{figure}
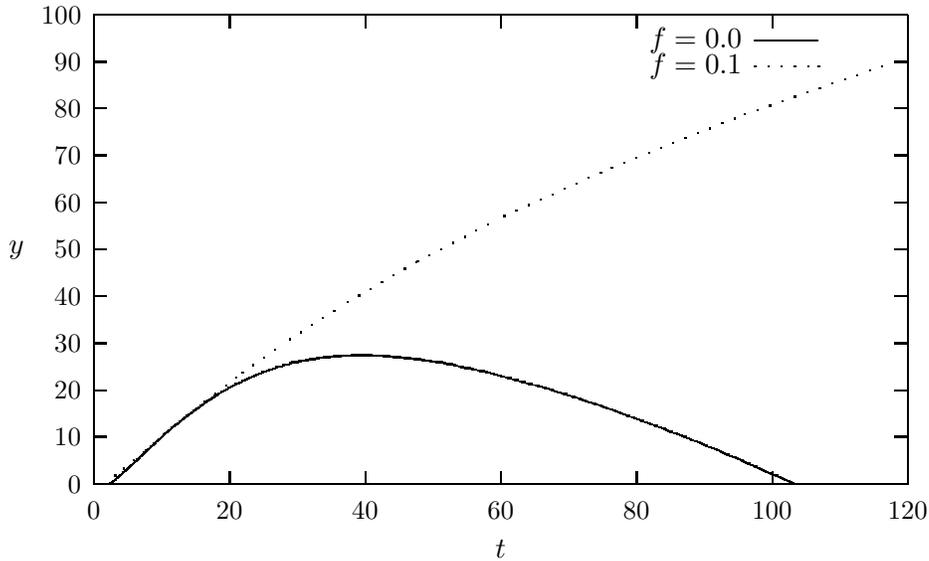

  \begin{center}
    \include{symmetry}
    \caption{Graph of $y$ against coördinate time for a geodesic in a
      cosmological braneworld with $\dot{y}_0 = 0.975$. The returning
      geodesic is in a $Z_2$-symmetric braneworld, while the one
      tending towards the horizon has $f=0.1$. The geodesics were set
      off at $a_0 = 0.5$}
    \label{fig:symmetry}
  \end{center}
\end{figure}

\begin{figure}
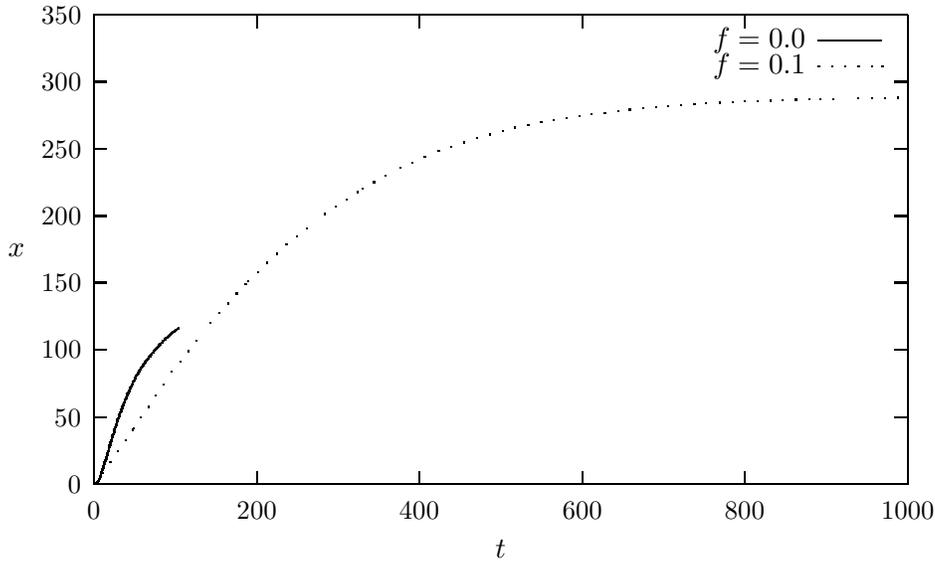

  \begin{center}
    \include{horizon}
    \caption{Graph of $x$ against coördinate time for a geodesic in a
      cosmological braneworld with $\dot{y}_0 = 0.975$. The geodesics
      are the same as in Figure \ref{fig:symmetry}. Note that the
      $f=0.1$ geodesic asymptotically tends to a maximal $x =
      x(\tau_H)$, where $\tau_H$ is the value of the affine parameter
      when the geodesic crosses the horizon.}
    \label{fig:horizon}
  \end{center}
\end{figure}

\begin{figure}
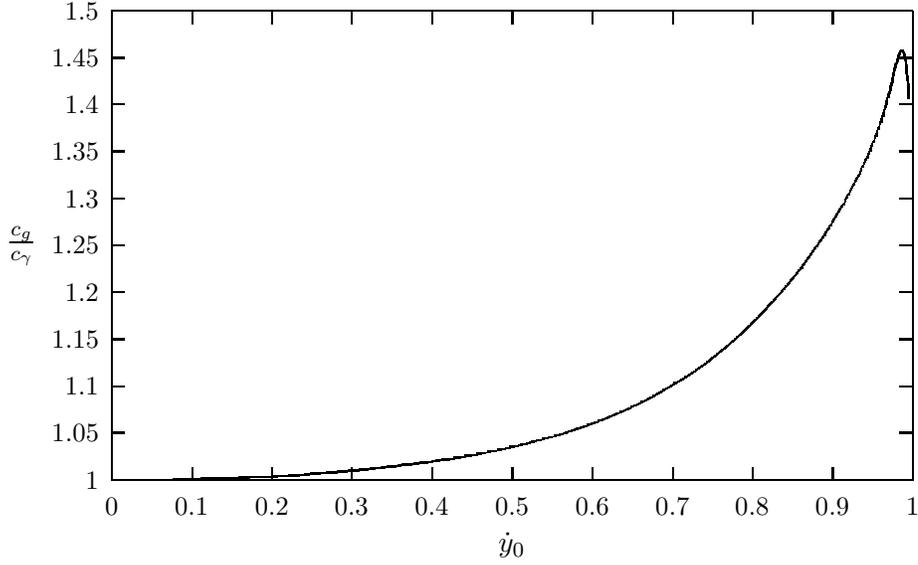

  \begin{center}
    \include{superluminal.f=0.a=0.5}
    \caption{The ratio of gravitational to electromagnetic speeds, as
      a function of the initial component of velocity in the fifth
      dimension, in a $Z_2$-symmetric braneworld. $a_0(t_0) = 0.5$.}
    \label{fig:normal}
  \end{center}
\end{figure}

\begin{figure}
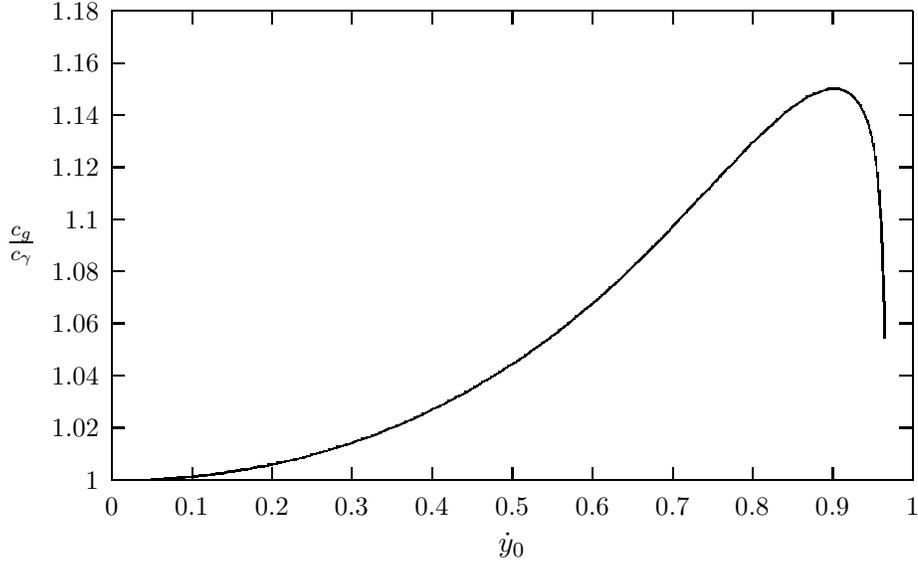

  \begin{center}
    \include{superluminal.f=0.1.a=0.5}
    \caption{The ratio of gravitational to electromagnetic speeds, as
      a function of the initial component of velocity in the fifth
      dimension, in a non-$Z_2$-symmetric braneworld ($f = 0.1$). Note
      firstly that the maximum has moved leftwards, and also that some
      geodesics from our samples no longer return.}
    \label{fig:asymmetry}
  \end{center}
\end{figure}

Consider figure \ref{fig:symmetry}, which shows null geodesics leaving
the brane at a time $t_0$ such that $a_0(t_0) = 0.5$, when the
$\rho^2$ term dominates in the Friedman equation\footnote{Though
  extensive tests have shown that there is no particular difference in
  between the régimes.}. In that figure, both geodesics have an initial
$\dot{y}$ of 0.975 (where the speed of light is 1); the difference is
that the dotted geodesic is in a braneworld with broken $Z_2$ symmetry
($f = 0.1$, in the notation of \cite{fpaper}). The graph has been cut
off so that the path of the returning geodesic is clear; the $f=0.1$
geodesic asymptotically (as $t$ tends to infinity) reaches the horizon
in the bulk. It is perhaps slightly clearer when viewed in conjunction
with figure \ref{fig:horizon}, which shows that 
when the geodesic in the $f=0.1$ world reaches $x$ of about 290 it is
at the horizon, at which point it will move with the horizon in the
$y$ direction.

Let us next examine figures \ref{fig:normal} and \ref{fig:asymmetry},
in which we have taken a number of geodesics as in figures
\ref{fig:symmetry} and \ref{fig:horizon}, and examined the
superluminosity of the returning signal; in other words, how much
faster than light, which is constrained to travel along the brane with
speed $\frac{1}{a_0(t)}$, is the true null geodesic.

The results are perhaps surprising at first sight. Conceptually there
are two separate effects happening here. One is that space is
`warped'; loosely, there is an $e^{\mu y}$ factor warping space in the
bulk, meaning that null geodesics can cover effectively much greater
distances. Thus, the farther into the bulk that the geodesic
penetrates, the more superluminal the geodesic will be. However, there
is a competing effect due to the presence of a horizon in the bulk;
the horizon is only a coördinate singularity, but since it is a
singularity in the physical brane coördinates it is nevertheless of
physical relevance, in that geodesics that reach the horizon never
return from the point of view of an observer on the brane.

Null geodesics that have too large a velocity component in the fifth
dimension will not be able to escape the horizon. Thus, at very 
high $\dot{y}_0$, the geodesics will not
return to the brane before the end of the universe. Null geodesics
with small initial $\dot{y}$, however, will be confined to the brane.

In between these two régimes, what will happen? At some critical
angle, the effect of the horizon will begin to dominate over the
effect of the warped spacetime, and consequently there will be a
maximum in the ratio of gravitational to electromagnetic effective
speeds.

As discussed in \cite{shortcuts}, the maximal superluminal effect will
be obtained when the hierarchy is the most pronounced
experimentally-allowed value, and the null geodesic is set off at the
earliest physical time possible (of the order of $M_5^{-1}$). However,
it is not because of any particularly stronger warping at this earlier
time, but rather because the geodesic can spend much `longer' in the
warped area before returning to the brane. Thus there is no
discrepancy in principle\footnote{Unfortunately it is not possible
without extensive code writing to numerically investigate this figure
in this framework, as the maximal superluminosity will be obtained
with an initial $\dot{y}$ of somewhere between $1-10^{-15}$ and 1.}
between the relatively low superluminal values shown here and the
figure of $10^3$ quoted in \cite{shortcuts}.

It appears from figures \ref{fig:normal} and \ref{fig:asymmetry} that
the effect of including asymmetry\footnote{We have shown here results
with positive $f$ only, as results with $f<0$ are physically similar.}
is to increase the horizon cross-section, for geodesics leaving at
equivalent times (taken as equal redshifts, or, what amounts to the
same thing, equal energy densities), and also to decrease the ratio of
the speeds even for geodesics that do not go anywhere near the
horizon.

\begin{figure}
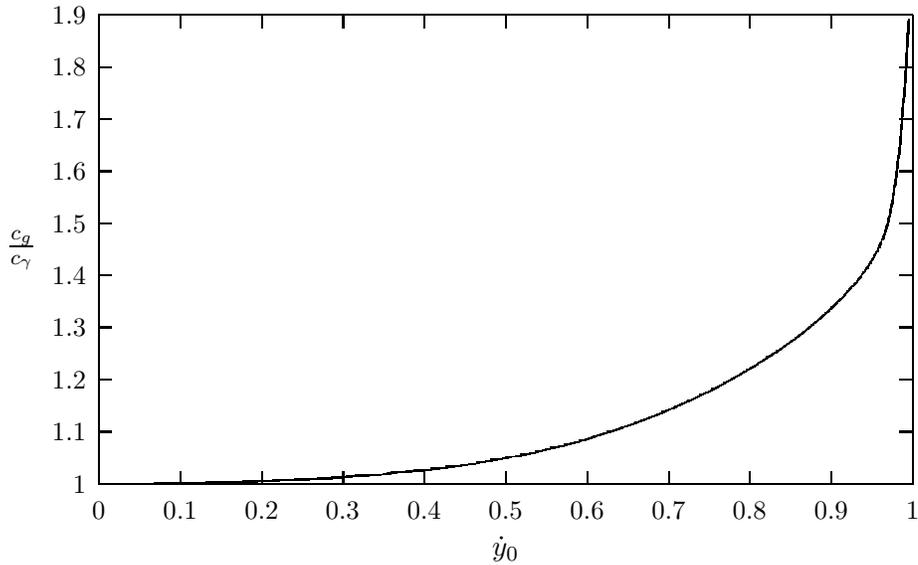

  \begin{center}
    \include{early}
    \caption{The ratio of gravitational to electromagnetic speeds, as
      a function of the initial component of velocity in the fifth
      dimension, in a $Z_2$-symmetric braneworld. $a_0(t_0) = 0.125$
      (earlier time than Figure \ref{fig:normal}).}
    \label{fig:early}
  \end{center}
\end{figure}

The effect of the start time on the superluminosity is such that when
the $\rho^2$ term dominates, at early times the superluminosity is
increased (see figure \ref{fig:early}). A corresponding late-time
graph is not shown, as the cross-section of the horizon has increased
to such an extent that almost all geodesics hit the horizon and do not
return, from which we note that the acausality in this model is
maximal at early time, with energy leakage from the brane at later
times.

\begin{figure}
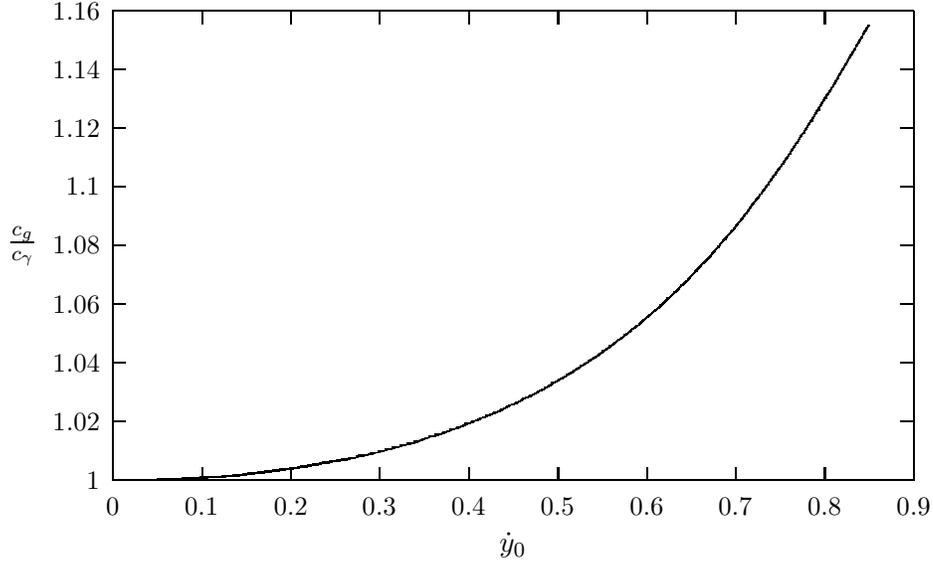

  \begin{center}
    \include{largec}
    \caption{The ratio of gravitational to electromagnetic speeds, as
      a function of the initial component of velocity in the fifth
      dimension, in a $Z_2$-symmetric braneworld. $w = 2.0$
      (same time as Figure \ref{fig:normal}). Note that some
      previously-returning geodesics no longer return.}
    \label{fig:largec}
  \end{center}
\end{figure}

Including dark radiation likewise dampens the superluminosity; figure
\ref{fig:largec} shows the effect of including a dark radiation term
$w=2$ in the calculation. Note that this value of $w$ is well outside
the nucleosynthesis bound \cite{bulkcosmo}, and is tested in our
investigation only so as to exaggerate any possible visible effect.

\section{The Two Brane Model}
\label{sec:twobrane}

It would seem from the above results, that the possibility of
null signals taking shortcuts through the bulk in the single brane
scenario would not solve the well known cosmological horizon problem
as the apparent speeds of such gravity signals are not significantly
greater than the speed of light signals confined to the brane. Due to this,
we now investigate an alternative suggestion made by~\cite{Freese}, 
which was that a 2-brane scenario could provide a solution to the horizon
problem. By investigating a variety of `toy' metrics, Chung and Freese showed
that apparently acausal signals could be sent between different points
on our brane, via the second hidden brane. These signals would, as
before, bring in to causal contact regions of our Universe that would
have otherwise been unable to communicate.

\begin{figure}
\center
\epsfig{file=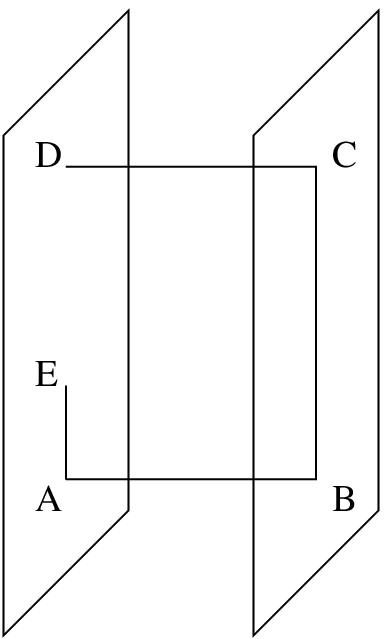,angle=0,scale=0.9}
\caption{A Null signal that remains on our brane goes from A to E in
time $t_f$, one which leaves the brane could travel between A, B, C
and then return to our brane at D in the same time as the first signal
reaches E.}
\label{figFTL}
\end{figure}

In order for this method to work, a metric of the following form
was assumed:
\begin{equation}\label{Fremetric}
\d s^2 \;\; = \;\; \d t^2 - e^{-2\mu y}a^2(t)\d x^2 - \d y^2 ,
\end{equation}
where $y$ corresponds to one extra spatial dimension. Our brane and
the hidden brane were assumed to be at $y=0$ and $y=\R$
respectively. The distance $D_{AE}$ travelled by a null signal on our brane
in time $t_f$ is then (see figure \ref{figFTL})
\begin{equation}
D_{AE} \;\;\;\; = \;\;\;\; \int^{t_f}_0 \frac{\d t}{a(t)} .
\end{equation}
However, if it is possible for a null signal to leave our brane,
interact with and therefore travel along the hidden brane, before
eventually returning to our world, we can then ask how far it would
have appeared to have travelled $D_{AD}$, in time $t_f$. If it takes a time
$t_c$ to cross between the branes then this distance would be given by
\begin{equation}
D_{AD} \;\;\;\; = \;\;\;\; e^{\mu \R} \int^{t_f-t_c}_{t_c}
\frac{\d t}{a(t)} .
\end{equation}
Therefore if $e^{\mu \R} \gg 1$ then $D_{AD} \gg D_{AE}$ and the
horizon problem will be solved. The major  problem with this
suggestion is that it involves metrics that are not realistic. It
relies upon there being a conformal factor that affects the spatial
part of the four-dimensional metric only. In typical Randall-Sundrum
Brane World models the metric is usually of the form
\begin{equation}\label{RSmetric1}
\d s^2 \;\;\; = \;\;\; e^{-2\mu y}\eta_{\mu\nu}\d x^{\mu}\d x^{\nu} -
\d y^2 ,
\end{equation}
and hence the horizon problem will remain unsolved.

\subsection{A Cosmologically Realistic Metric}

The above metrics (\ref{Fremetric}) and (\ref{RSmetric1}) are
obviously not realistic descriptions of our Universe; it is
therefore worthwhile to consider if the above proposed solution to the horizon
problem will work for a cosmological metric.
If $w$, $f$ and $k$ are set to zero, the solutions to Einstein's
equations given by (\ref{met1_form}) can be greatly simplified. Writing
$a(t,y)$ in terms of exponentials gives
\begin{equation}
a(t,y)^2 \;\;\; = \;\;\; \frac{a_0^2 e^{-2\mu y}}{4} 
                   \left[
                         \frac{\rho}{\sigma} e^{2\mu y} - 
                         (\frac{\rho}{\sigma}+2)
                   \right]^2,
\end{equation}
This can be further simplified to give
\begin{equation}\label{a_1}
a(t,y) \;\;\; = \;\;\; a_0\left(
                                \co - \eta_0 \si
                          \right),
\end{equation}
which can be used with equation (\ref{def_n}) to give
\begin{equation}\label{n_1}
n(t,y) \;\;\; = \;\;\;  \left(
                              \co - \tilde{\eta}_0 \si
                        \right),
\end{equation}
We have defined 
\begin{equation}
\eta_0 \;\; = \;\; 1+\frac{\rho}{\sigma}, \;\;\;\;\; 
\tilde{\eta}_0 \;\;= \;\;\eta_0 
+ \frac{\dot{\eta}_0}{H_0}.
\end{equation}
Here, $H_0$ is the Hubble constant on the brane at $y=0$, given by
equation \ref{Freid1} and can now be
written in the form
\begin{equation}\label{H_0}
H_0^2 \;\;\; = \;\;\; \frac{\kappa^4}{36} \rho_0^2 - \frac{\Lambda}{6} 
\;\;\; = \;\;\; \mu^2 \left(\eta_0^2 -1\right) .
\end{equation}
In order for there to be a significant difference in the relative
`speeds' of null signals travelling along the two branes, we require
that $n/a$ is larger for some $y$ than it is for $y=0$, i.e. that
\begin{equation}
\frac{n(t,y)}{a(t,y)} \gg \frac{n(t,0)}{a(t,0)} \;\; \Leftrightarrow \;\;
\frac{\co - \tilde{\eta}_0 \si}{a_0(\co - \eta_0\si)} \gg \frac{1}{a_0}.
\end{equation}
An acceptable equation of state on our brane demands that $\rho
\propto a_0^{-q}$ and this implies that $\tilde{\eta}_0 = \eta_0 -q
\rho/\sigma$ and therefore that $\tilde{\eta}_0 < \eta_0$. Because of
this the ratio $n/a$ will diverge as $y\rightarrow y_h$, where $y_h$
is the position of the horizon defined by $a(t,y_h)=0$ and is given by
\begin{equation}
\tanh{\mu y_h} \;\;\; = \;\;\; \frac{1}{\eta_0}.
\end{equation}
This suggests that null signals may travel along the second brane at
much greater `speeds' than on our brane, provided the second brane is
close to the horizon. Unfortunately, to correctly calculate the
difference in speeds, we need to know the motion of the second brane
with respect to the one at $y=0$.

\subsection{General Behaviour of the Inter-brane Distance}

In this section we use a simple non-perturbative method to derive the general 
equation of motion for the inter-brane distance otherwise known as the
cosmological radion and denoted by $\R(t)$, which was first derived 
by~\cite{bin3}. For alternative approaches to this topic 
see~\cite{radionwave,stabmodulus,radionds,locallylocal}.

Noting first that if  the second brane follows a trajectory given by
$y=\R(t)$, then the induced 4-dimen\-sional metric on the second brane
is given by
\begin{eqnarray}
\d s^2 \;\;\; &=& \;\;\; -\left[
                        n^2(t,\R(t))- \dot{\R}^2
                        \right]\d t^2
                 + a^2(t,\R(t))\d x^2  \\
              &=& \;\;\; -\d \tau^2 + a_2^2(\tau) \d x^2 ,
\end{eqnarray}
where $\tau$ has been defined as the proper time as seen by an
observer on the second brane. The expansion rate of 
the second brane as seen by an observer on our brane is simply
\begin{equation}
H_2(t) \;\;\; = \;\;\; \frac{1}{a_2}\frac{\d a_2}{\d t} 
       \;\;\; = \;\;\; \left(
                       \frac{\dot{a}}{a} + \frac{a'}{a}\dot{\R}
                       \right)_2,
\end{equation}
and as seen by an observer on the second brane itself
\begin{eqnarray}\label{H_2}
\mathcal{H}_2(\tau) \;\;\; &=& \;\;\; \frac{1}{a_2}
\frac{\d a_2}{\d \tau} \;\;\; = \;\;\; H_2(t) \frac{\d t}{\d \tau} \\
         &=& \;\;\;
                  \left(     
                       \frac{\dot{a}}{a} + \frac{a'}{a}\dot{\R}     
                  \right)                  
                  \left( 
                       n^2- \dot{\R}^2
                  \right)^{-1/2} .             
\end{eqnarray}
This equation for $\HU_2$ is important as it relates the expansion
rate of the second brane (with respect to proper time) to $\R$ and 
$\dot{\R}$. It does not seem so useful at first, as calculating $\HU_2$
could be difficult; however, we know that the brane world Friedman
equation given by (\ref{H_0}) is derived from a purely local analysis
and that should we have chosen the second brane to be stationary and
at $y=0$, we would have derived the equivalent Friedman equation with
$\rho$ replaced by $\rho_2$. This means that $\HU_2(\tau)$ must have
the form
\begin{equation}\label{H_3}
\HU_2^2(\tau) \;\;\; = \;\;\; 
                   \frac{\kappa^4}{36} \rho_2^2 - \frac{\Lambda}{6} 
\;\;\; = \;\;\; \mu^2 \left(\eta_2^2 -1 \right).
\end{equation}
Where we have defined $\eta_2 = \rho_2/\sigma$.
Equation (\ref{H_3}) ensures that the second brane is $Z_2$ symmetric and
that it evolves according to the junction conditions of the
extrinsic curvature tensor, just as equation (\ref{H_0}) ensures
similar behaviour for our
brane. At this point we use the following identity\footnote{Note that
although the derivation is done here for $f=k=w=0$, this method
generalizes to braneworlds with non-zero values of these parameters
with the use of the appropriate Friedman equation for equation \ref{H_3}
and identity for equation \ref{eq:identity}.}
which
is obtained from the Einstein equations and is derived in \cite{bulkcosmo}:
\begin{equation}
\label{eq:identity}
\left(\frac{\dot{a}}{na} \right)^2 \;\; = \;\;
         \frac{a'^2}{a^2} \; - \; \mu^2,
\end{equation}
giving
\begin{equation}\label{H_4}
\HU_2^2 \;\;\; = \;\;\; 
     \left(\frac{\dot{a}}{na} \right)^2 - \frac{a'^2}{a^2} 
       +\mu^2 \eta_2^2.
\end{equation}
Substituting (\ref{H_4}) into (\ref{H_2}) and rearranging finally gives
the following first order equation for $\dot{\R}$:
\begin{equation}\label{R1}
\dot{\R} \;\;\; = \;\;\;
        n \left( 
               -\frac{a'\dot{a}}{a^2n} \; \pm \;
               \mu^2 \eta_2 \sqrt{\eta_2^2 - 1} 
          \right) 
          \left( 
                \frac{\dot{a}^2}{n^2a^2} + \mu^2 \eta_2^2 
          \right)^{-1}.
\end{equation}
This can be rewritten in the form found by~\cite{bin3}
\begin{equation}\label{R2}
\frac{a'}{a} \; + \; \frac{\dot{a}}{n^2a} \dot{\R} \;\; = \;\; 
     \mu \eta_2 \left(1 - \frac{\dot{\R}}{n^2}\right)^{1/2}.
\end{equation}
In order to investigate acausal signals we will be interested in a
stationary second brane, and therefore need an equation for
$\ddot{\R}$.~\cite{bin3} have shown that by differentiating (\ref{R2})
with respect to time, considering the three-dimensional symmetries of
the bulk energy momentum tensor, and the $Z_2$-symmetry of the branes,
it is possible to derive the following equation for $\ddot{\R}$
\begin{equation}\label{Rdd}
\frac{\ddot{\R}}{n^2} \;+\; \frac{n'}{n}
           \left(
                 1-2\frac{\dot{\R}^2}{n^2}
           \right)
 \;-\; \frac{\dot{n}}{n} \frac{\dot{\R}}{n^2} \;\; = \;\;
        -(2\mu \eta_2 + 3\mu p_2)
           \left(
                 1-\frac{\dot{\R}^2}{n^2}
           \right)^{3/2}.
\end{equation} 
Equations (\ref{R1}) and (\ref{Rdd}) govern the evolution of $\R$ and
will be used in the next two sections to evaluate the difference in the speeds
of null signals travelling along each brane.

\subsection{Acausal Signals: Stationary Branes}

The simplest solutions to equations (\ref{R1}) and (\ref{Rdd}) are
those describing a stationary second brane; in this section we will
investigate the appropriate conditions and the ratio of distances
travelled on each brane in a certain time interval.
Setting $\dot{\R} = \ddot{\R} = 0$ in (\ref{R2}) and (\ref{Rdd})
trivially gives the well known conditions on $\eta_2$ and $p_2 \equiv P_2/\sigma$ which 
ensure that the second brane does not move:
\begin{eqnarray}\label{stab1}
\eta_2 \;\;\; &=& \;\;\;
               \frac{\sir - \eta_0\cor}{\cor - \eta_0\sir} , \\
p_2 \;\;\; &=& \;\;\;
        -\frac{1}{3} 
         \frac{\sir - \tilde{\eta}_0\cor}{\cor - \tilde{\eta}_0\sir}
        -\frac{2}{3} \eta_2 .
\end{eqnarray}
Note that if we demand standard cosmology on our brane at late times
then this requires that both $\eta_0$ and $\tilde{\eta}_0$ tend to
1 (not zero as~\cite{bin3} have suggested). This results in the equation 
of state of the second brane becoming $-\eta_2 = p_2 = 1$ which 
is equivalent to $-\rho_2 = P_2 = 6\mu/\kappa^2$, the conditions in
the original Randall-Sundrum two brane scenario. 
Assuming the above form for the energy and pressure densities on the
second brane, we now just have to evaluate the distances travelled by
null signals between times $t_1$ and $t_2$ on either brane:
\begin{equation}\label{D1D2}
 D_1 \;\; = \;\; \int_{t_1}^{t_2} \frac{1}{a_0} \d t, \;\;\;\;\;\;
 D_2 \;\; = \;\; \int_{t_1}^{t_2} 
      \frac{\sqrt{
                 n^2(t,\R(t))-\dot{\R}^2
                  }}
           {a^2(t,\R(t))}
           \d t,
\end{equation}
however since we have assumed a stationary second brane, $D_2$
simplifies to become
\begin{equation}
D_2 \;\; = \;\; 
              \int_{t_1}^{t_2} \frac{n(t,\R(t))}{a(t,\R(t))} \d t
    \;\; = \;\; 
              \int_{t_1}^{t_2} \frac{\cor-\tilde{\eta}_0\sir}
                               {a_0(\cor-\eta_0\sir)} \d t.
\end{equation}
As mentioned before, a standard equation of state on our brane leads
to $\tilde{\eta}_0$ having the form $\tilde{\eta}_0 = \eta_0
-q\rho/\sigma$ and therefore,
\begin{eqnarray}
D_2 \;\;&=& \;\;
            \int_{t_1}^{t_2} 
            \left[
                  \frac{1}{a_0} \; + \;
                  \frac{(q\rho/\sigma)\sir}{a_0(\cor-(1+\rho/\sigma)\sir)}
            \right]\d t \\
        &=& \;\; D_1 \;\; + \;\; \Delta D.
\end{eqnarray}
Rewriting the Friedman equation for our brane in terms of $\rho/\sigma$
gives $H_0^2 = \mu^2 q^2( (\rho/\sigma)^2 +2\rho/\sigma)$. This, and
the fact that $\rho = \lambda/a_0^q$ allows the $\Delta D$ integral to
be converted into the following form:
\begin{equation}
\Delta D \;\; = \;\;
             \frac{A(\R)}{\mu(\sigma \lambda)^{1/q}}
             \int_{u_1}^{u_2}
             \frac{\d u}{u^{1/q} \sqrt{2u+1}(u-A(\R))}
\end{equation}
where $u=\sigma/\rho$ and all the information on the second brane's
position has been grouped into $A(\R)= \sir/(\cor-\sir)$. The integral
in (\ref{D1D2}) can be similarly converted, and we then obtain the
ratio of the extra distance travelled on the second brane, divided by
the distance travelled on our brane in the same time,
\begin{equation}\label{int1}
\frac{\Delta D}{D_1} \;\; = \;\; 
             qA(\R)
             \left(
                   \int_{u_1}^{u_2}
                   \frac{\d u}{u^{1/q} \sqrt{2u+1}(u-A(\R))}
             \right)
              \left(
                    \int_{u_1}^{u_2}
                    \frac{\d u}{u^{1/q} \sqrt{2u+1}}
              \right)^{-1}.
\end{equation}
We now need to determine the possible limits on $u_1$ and
$u_2$. Obviously the contribution of the integral to $\Delta D$ for
values of $u$ greater than $(q+1)A(\R)$ will be less than the
corresponding contribution to $D_1$ and so we choose the upper limit
on $u_2$ to be $(q+1)A(\R)$, as a larger $u_2$ is of no
interest. The minimum possible value of $u_1$ is slightly harder to
find as it is constrained by requiring that the energy densities of
both branes remain below the 5-dimensional Planck mass limit
\begin{equation}
\rho \; \le \; M_5^4  \;\;\; \Rightarrow \;\;\; 
       u \; = \; \frac{\sigma}{\rho} \;\;\; 
              \le \;\;\; \frac{6M_5^2}{M_4^2} \; 
              = \;    \frac{6}{M} ,
\end{equation}
where we have defined $M$ to be the dimensionless ratio of the squares
of the four- and five-dimensional Planck masses $M \equiv M_4^2 /
M_5^2$. Demanding a similar constraint on $\rho_2$ and using the
stability condition given by (\ref{stab1}), results in the following
condition
\begin{eqnarray}
|\rho_2| \; \le \; M_5^4  \;\;\; 
       &\Rightarrow& \;\;\; |\eta_2| \; = \; \frac{|\rho_2|}{\sigma} 
                     \;\;\; \le \;\;\; \frac{M}{6} \\
       &\Rightarrow& \;\;\; 
                    \frac{\eta_0 \cor - \sir}{\cor -\eta_0 \sir}
                    \;\; \le \;\; \frac{M}{6}.                                 
\end{eqnarray}
This can be rearranged and expressed in terms of $u$ using the fact
that $\eta_0 = 1+1/u$, to give the constraint on $u$ that ensures that
$|\rho_2|$ is always less than $M_5^4$:
\begin{equation}\label{ulimit}
u \;\;\; \ge \;\;\; 
             \frac{ M \tar + 6}
                  {(M-6)(1-\tar)} .
\end{equation}
In the next two sections we will use this constraint to determine
$u_1$ which corresponds to the earliest possible time that a null
signal can set off along the second brane. We examine both the near
brane ($\mu\R \ll 1$) and the far brane ($\mu\R \gg 1$) limits.

\subsection{Near Brane Limit}

We will now evaluate the integral expression for $\Delta D /D_1$ given
by (\ref{int1}), in the small $\mu\R$ limit.
The minimum possible value of the inter-brane distance $\R$ will be the
inverse of the 5D Planck mass: $\R \ge 1/M_5$ and for a cosmologically
realistic brane $\mu$ is given by $\mu = M_5^3/M_4^2$, leading to the
relation
\begin{equation}
\mu \R \;\;\; \ge \;\;\; \frac{M_5^2}{M_4^2} \;\;\; = \;\;\;
\frac{1}{M} .
\end{equation}
If we therefore assume $1/M \le \mu\R \ll 1$, we then find that in
this limit $A(\R)\simeq \mu \R$ and therefore that $u_1<u_2 =
(q+1)A(\R) \ll 1$. Equation (\ref{int1}) can then be approximated by
\begin{equation}\label{int2}
\frac{\Delta D}{D_1} \;\; \simeq \;\; 
             q A(\R)
             \left(
                   \int_{u_1}^{(q+1)A(\R)}
                   \frac{\d u}{u^{1/q} (u-A(\R))}
             \right)
              \left(
                    \int_{u_1}^{(q+1)A(\R)}
                    \frac{\d u}{u^{1/q}}
              \right)^{-1} .
\end{equation}
If we now assume that our brane is undergoing radiation dominance by
setting $q=4$, the first integral on the right hand side of
(\ref{int2}) can be integrated to give
\begin{eqnarray*}\label{int3}
\lefteqn{ A^{\frac{1}{4}}  \int_{u_1}^{u_2}  \frac{\d u}{u^{1/4}
     (u-A(\R))} \; = }    \\ 
     & &  2\arctan\left[
                        \left(\frac{u_2}{A}\right)^{\frac{1}{4}}       
                  \right] \; - \; 
          2\arctan\left[
                        \left(\frac{u_1}{A}\right)^{\frac{1}{4}}      
                  \right] \; + \;
         \ln\left[
                  \frac{\left(\frac{u_2}{A}\right)^{\frac{1}{4}} - 1}
                       {\left(\frac{u_1}{A}\right)^{\frac{1}{4}} - 1} \;.\;
                  \frac{\left(\frac{u_1}{A}\right)^{\frac{1}{4}} + 1}
                       {\left(\frac{u_2}{A}\right)^{\frac{1}{4}} + 1}
            \right],
\end{eqnarray*}
and the second integral is trivially given by
\begin{equation}
\int_{u_1}^{u_2}
             \frac{\d u}{u^{1/4}} \;\;\; = \;\;\;
           \frac{4}{3} ( u_2^{\frac{3}{4}} - u_1^{\frac{3}{4}} ).
\end{equation} 
From above we take $u_2 = (q+1)A(\R)= 5A(\R)$, and therefore we just
need to evaluate $u_1/A(\R)$ in the small $\mu \R$ limit using equation
(\ref{ulimit})
\begin{equation}
      \frac{u_1}{A(R)}   \;\; = \;\; 
      \frac{M \tar \;+\; 6}
          {(M \;-\; 6)\tar}  \;\; \simeq \;\;
               1 \;+\; \frac{3}{2M \mu \R}.
\end{equation}
Replacing the values for $u_1$ and $u_2$ into the expression for
$\Delta D/D_1$ and ignoring all subdominant terms, finally gives the
ratio of the maximum extra distance travelled by a null signal on the second
brane compared to the first in the near brane limit as
\begin{equation}\label{D1near}
\frac{\Delta D}{D_1} \;\;\; \simeq \;\;\; 1.28  \ln (M \mu \R).
\end{equation}
The value of $M_5$ is constrained by nucleosynthesis such that $M_5 \ge
30$TeV. This leads to the corresponding constraint on the
dimensionless ratio $M$ to be
\begin{equation}\label{M1}
M \;\;\; = \;\;\; \frac{M_4^2}{M_5^2} \;\;\; < \;\;\; 10^{29}. 
\end{equation}
In order for the above approximations to be valid $\mu\R$ must satisfy
$\mu\R < 10^{-2}$, therefore (\ref{D1near}) becomes
\begin{equation}\label{D2near}
\frac{\Delta D}{D_1} \;\;\; \simeq \;\;\; 80.
\end{equation}
This shows that any seemingly acausal effects due to signals travelling
on the second brane as opposed to our brane are too small to solve the
cosmological horizon problem in the ($\R \ll 1/\mu$) limit. Another
important point is that $\Delta D$ is only significantly greater than
$D_1$ for a very brief time, shortly after the big bang. If the signals
were to travel for longer periods, say until $u_2 = 1$ then the
distance ratio
would become much less than 1: $\Delta D/D_1 \simeq \mu\R \ln(M\mu\R)$ 
as can be seen from (\ref{int2}).

\subsection{Far Brane Limit}

In this Section we investigate the behaviour of null signals
travelling on the second brane when the inter-brane distance is large
($\R \gg 1/\mu$). The fact that the second brane has to be closer to our brane
than the horizon implies that we will be examining signals that
are travelling at late times with respect to our Universe, as opposed
to the previous Section where the only interesting situations occurred 
at early times. 

In the $\R \gg 1/\mu$ limit, the function $A(\R)$ becomes very large
and can be approximated by
\begin{equation}
A(\R) \;\;\; = \;\;\; \frac{\tar}{1 - \tar} \;\;\; \simeq \;\;\;
                      \frac{1}{2} e^{2\mu \R}.
\end{equation}
As before, we will require the ratios $u_2/A(\R) = (q+1)$ and
$u_1/A(\R)$ which is obtained from equation (\ref{ulimit}) and is
approximately given by 
\begin{equation}\label{u1large}
      \frac{u_1}{A(R)}   \;\; = \;\; 
      \frac{M \tar \;+\; 6}
          {(M \;-\; 6)\tar}  \;\; \simeq \;\;
               1 \;+\; \frac{12}{M}.
\end{equation}
Therefore both $u_1$ and $u_2$ are exponentially large and so the
expression for $\Delta D/D_1$ given by equation (\ref{int1}) can be 
simplified in the large $\R$ limit to
\begin{equation}\label{int4}
\frac{\Delta D}{D_1} \;\; \simeq \;\; 
             q A(\R)
             \left(
                   \int_{(1 + \frac{12}{M})A(\R)}^{(q+1)A(\R)}
                   \frac{\d u}{u^{\frac{1}{q} + \frac{1}{2}} (u-A(\R))}
             \right)
              \left(
                    \int_{(1 + \frac{12}{M})A(\R)}^{(q+1)A(\R)}
                    \frac{\d u}{u^{\frac{1}{q} + \frac{1}{2}}}
              \right)^{-1}.
\end{equation}
Assuming radiation dominance, the first integral on the right hand
side of (\ref{int4}) can be solved to give
\begin{eqnarray*}\label{int5}
\lefteqn{ A^{\frac{3}{4}}  \int_{u_1}^{u_2}  \frac{\d u}{u^{3/4}
     (u-A(\R))} \; = }    \\ 
     & & -2\arctan\left[
                        \left(\frac{u_2}{A}\right)^{\frac{1}{4}}       
                  \right] \; + \; 
          2\arctan\left[
                        \left(\frac{u_1}{A}\right)^{\frac{1}{4}}      
                  \right] \; + \;
         \ln\left[
                  \frac{\left(\frac{u_2}{A}\right)^{\frac{1}{4}} - 1}
                       {\left(\frac{u_1}{A}\right)^{\frac{1}{4}} - 1} \;.\;
                  \frac{\left(\frac{u_1}{A}\right)^{\frac{1}{4}} + 1}
                       {\left(\frac{u_2}{A}\right)^{\frac{1}{4}} + 1}
            \right],
\end{eqnarray*}
which when combined with (\ref{int4}) and (\ref{u1large}), and after
all subdominant terms are neglected, results in the following
expression for the extra distance travelled on the second brane in the
far brane limit
\begin{equation}\label{D1far}
\frac{\Delta D}{D_1} \;\;\; \simeq \;\;\; 2 \ln (M).
\end{equation}
Note that if instead of radiation domination we
had chosen matter domination and set $q=3$, the leading term given in
equation (\ref{D1far}) would not have been altered. Replacing into 
(\ref{D1far}) the maximum possible value of $M \simeq
10^{29}$ leads to the final result
\begin{equation}\label{D2far}
\frac{\Delta D}{D_1} \;\;\; \simeq \;\;\; 130.
\end{equation}

Unlike when $\R \ll 1/\mu $, here $\Delta D$ is significantly greater
than $D_1$ for an extended period of time. Solving the Friedman
equation gives the late time relation between $u$ and the cosmic time
$t$ as measured on our brane to be $u \simeq \mu^2 q^2 t^2/2$. This
implies that $\Delta D > D_1$ during a time interval of $\Delta t
\simeq (M_4^2/2M_5^3)\exp(\mu\R) = 10^{27}\exp(\mu \R)$TeV$^{-1}$.
However, as can be seen from equation \ref{int4}, if limits for the
integral for $\Delta D/D_1$ are such that $u_2 \gg u_1$, then any
acausal effect becomes negligible. Taking $u_1$ to be just before
recombination and $u_2$ at the present day leads to $u_1 \sim 10^{12}$
and $u_2 \sim 10^{21}$, and therefore the horizon problem cannot be
solved in this manner, though this effect may have important
consequences in other areas of cosmology such as structure
formation. One should note that the effectiveness of the acausality
would also be lessened by the time taken for signals to cross between
the branes.

\subsection{Acausal Signals: Moving Branes}

Up till now we have been examining the possibility of acausal signals
travelling along stationary branes. It is however interesting to
consider the effect of allowing the branes to move. As will be shown, 
for a general
cosmological scenario the equation of motion of the second brane is
very difficult to solve; various qualitative features can, however, be
investigated and will be discussed in this section. We start with the
equation for the time dependent inter-brane spacing $\R (t)$, derived
previously (\ref{R2})
\begin{equation}\label{R5}
\frac{\dot{\R}}{n} \;\;\; = \;\;\;
          \left( 
               -\frac{a'\dot{a}}{a^2n} \; + \;
               \mu^2 \eta_2 \sqrt{\eta_2^2 - 1} 
          \right) 
          \left( 
                \frac{\dot{a}^2}{n^2a^2} + \mu^2 \eta_2^2 
          \right)^{-1}.
\end{equation} 
This can be rewritten using equations (\ref{H_3}), (\ref{H_4}),
(\ref{a_1}) and (\ref{n_1}) as
\begin{equation}\label{R6}
\frac{\dot{\R}}{n} \;\; = \;\;
                \frac{ H_0(\eta_0 \cor - \sir) + \mu\eta_2\HU_2
                      (\cor - \eta_0\sir)^2}
                     {(\eta_0 \cor - \sir)^2 + \HU^2_2 
                      (\cor - \eta_0\sir)^2}.
\end{equation}
For the trivial case where $\eta_2$ is constant and therefore 
$\HU_2=0$, equation (\ref{R6}) can easily be solved numerically as was
done in~\cite{bin3}. For a cosmologically realistic case where $\eta_2$ is
time dependent, the solution is less easily obtained. If, for
example, we assume a standard equation of state on the second brane:
$P_2 = \omega_2 \rho_2$ which provided $\omega_2>-1$, implies that 
$\rho_2 \propto 1/a_2^{q_2}$, we can then solve the Friedman equation
\begin{equation}
\HU_2^2 \; = \; \frac{1}{a_2^2}\left(
                     \frac{\d a_2}{\d\tau} \right)^2 \; = \;
                \frac{1}{q_2^2 \eta_2^2}\left(
                      \frac{\d\eta_2^2}{\d\tau}\right)^2 \; = \;
                \mu^2(\eta_2^2-1) ,
\end{equation}
to give $\eta_2 = -1/\sin(q_2 \mu \tau)$ and $\HU_2 = \mu^2/
\tan(q_2\mu \tau)$. However, to solve equation (\ref{R6}) we need to
know $\eta_2$ and $\HU_2$ in terms of $t$: our cosmic time, which is
related to the second branes cosmic time $\tau$ by
\begin{equation}
\d\tau \;\;\; = \;\;\; \sqrt{n^2 - \dot{\R}^2} \; \d t.
\end{equation}
This makes it much more difficult to solve for $\dot{\R}$ and due 
to this added level of complication we leave a more general study of 
(\ref{R6}) to future work. We now examine the situation
qualitatively and argue that acausality in the moving brane model
is not appreciably greater than in the static case.

From equation (\ref{R5}) it can be seen immediately that since
$a'<0$ for $0<R<y_h$, then if 
$\eta_2 >0$ then $\dot{\R} >0$ for
all $R$ and the second brane will move away from our brane and freeze
out at the horizon, which has a time dependent position given by $y_h
= \arctan(1/\eta_0)$. If $\eta_2 < 0$, then $\dot{\R} =0 $ only when
\begin{equation}
\mu \eta_2 \;\;\; = \;\;\; \frac{a'(t,\R)}{a(t,\R)}.
\end{equation}

Requiring also that $\ddot{\R}=0$ gives the conditions for stationary
branes as discussed previously. The stability of a stationary second
brane has been investigated in~\cite{bin3} where it was shown that if
both branes are de~Sitter the equilibrium position is unstable, while
if both branes are anti-de~Sitter then it is stable. For our purposes,
we only need to examine the case where the second brane moves in the
positive $y$ direction and freezes out at the horizon, as the other
possibilities are that the second brane approaches our brane and
collides with it which is physically unacceptable, or that the second
brane is stable: the case which has been examined previously.

If the second brane does freeze out, then $\R \rightarrow y_h$ and  
$\dot{\R} \rightarrow n(t,R)$ as can be checked from equation
(\ref{R5}). The expression for the distance
travelled on the second moving brane is
\begin{equation}
 D_2 \;\; = \;\; \int_{t_1}^{t_2} 
      \frac{\sqrt{
                 n^2(t,\R(t))-\dot{\R}^2
                  }}
           {a^2(t,\R(t))}
           \d t,
\end{equation}
which will therefore tend to zero as $\dot{\R} \rightarrow
n(t,R)$. Another problem is the distance that signals have to travel
to return. In order for there to be a significant acausal effect, the
null signal has to be travelling on the second brane for a substantial
amount of time and therefore would be a large distance away from our
brane when starting to return. The combination of these two effects:
the distance travelled by a null signal on the second brane frozen at 
the horizon tending to zero, and the large distance returning signals
would have to travel, suggest that little or no acausal effects would be
observed on our brane.

\section{Discussion}
\label{sec:conclusions}

We have investigated the behaviour of null geodesics in several
five-dimensional brane world scenarios. In the single brane case it
was shown that apparent causality violations caused by such signals
taking shortcuts through the bulk were small. The ratio of the speed
of five-dimensional gravity signals to the speed of four-dimensional
light $c_g / c_{\gamma}$ was found to be in general not more than
two. The effects of introducing a non-zero Weyl tensor component and of
relaxing the mirror (or $Z_2$) symmetry were examined and shown in
general to decrease the observed acausality.

It was found that during early times in the Universe many geodesics
would return to our brane, as opposed to late times when most of the
geodesics leaving our brane freeze out at the horizon.
This is understandable, as at late times ($\rho \rightarrow 0$) the
cosmological metrics that were considered tend to the standard
Randall-Sundrum metric which, as previously discussed, results in
geodesics not returning. The behaviour of these geodesics could be
used in conjunction with the temperature dependent production rate of
gravitons to exactly determine the amount of energy lost to the bulk
throughout the history of the Universe as is discussed
in~\cite{bulkbh}.

Analysis of the two brane scenario (with $w=f=k=0$) shows that signals
can in some situations travel along the second brane significantly
faster than along our brane. This effect, however, would only last for
a certain time depending on the position of the second brane. In the
near brane limit ($\mu \R \ll 1$) this period is very brief and any
acausal effects would become negligible when longer time intervals are
considered. In the far brane limit ($\mu \R \gg 1$) however, the
period of interest can last for $\Delta t \simeq 10^{27}\exp(\mu
\R)$TeV$^{-1}$and it was found that the apparent speeds of null
signals were approximately 130 times faster on the second brane than
on the first. It should be noted that a non-zero $w$ and $f$ would
lessen the above effect as the crucial ratio $n(t,\R)/a(t,\R)$ is
largest only when $w=f=0$. Another factor that would detract from the
effectiveness of this mechanism in the far brane limit is the time
taken for signals to travel between the two branes.

In neither one nor two brane scenarios are the possible acausal
effects significant enough to enable us to solve the cosmological
horizon problem. The increase in the gravitational particle horizon
could, however, have important effects on other areas of cosmology
such as the initial conditions of inflation, structure formation and
the production of topological defects. It would also technically be
possible to measure the time delay between the detection of
gravitational waves and light waves from a particular cosmic event;
however, we believe that this would require significant advances in
gravitational wave detector technology.

There is also the possibility that a higher dimensional model could
produce a large enough acausal effect to solve the horizon
problem~\cite{shap6D,gogber6D,chod1,cohandkap,greg1,itsaso,univasp,critcoscon}.
Unfortunately, we are currently unaware of any metrics in ($d+1$)
dimensions with $d>4$ where acausal signals are possible. For example
there have been several six-dimensional `brane world' models proposed
that have a metric Ansatz of the general form
\begin{equation}
\d s^2 \;\;\; = \;\;\;
     \phi^2(x^i) \eta_{\alpha\beta}(x^{\nu}) \d x^{\alpha} \d
     x^{\beta}  \;\; + \;\;
     g_{jk}(x^i) \d x^j \d x^k,  
\end{equation}
where $\alpha, \beta$, and $\nu$ run across time and normal space
dimensions 
(0 to 3) and $i$ and $j$ run across the two extra spatial dimensions
(5 to 6). It can be seen that no acausal effects are possible since
both the time and the three-dimensional spatial components of the
metric have the same dependence on the extra dimensions, unlike the
five-dimensional case where $n(t,y) \not= a(t,y)$. A fully
cosmological six-dimensional metric could however overcome this problem.

\acknowledgments

This work is supported in part by PPARC.

\bibliography{brane}
\bibliographystyle{JHEP}

\end{document}